 \documentstyle[multicol,aps,prl,eqsecnum]{revtex}
 \renewcommand{\narrowtext}{\begin{multicols}{2} \global\columnwidth20.5pc}
 \renewcommand{\widetext}{\end{multicols} \global\columnwidth42.5pc}  
\input{epsf.tex}
\def\inseps#1#2{\def\epsfsize##1##2{#2##1} \centerline{\epsfbox{#1}}}

\def\top#1{\vskip #1\begin{picture}(290,80)(80,500)\thinlines \put(
65,500){\line( 1, 0){255}}\put(320,500){\line( 0, 1){
5}}\end{picture}}
\def\bottom#1{\vskip #1\begin{picture}(290,80)(80,500)\thinlines \put(
330,500){\line( 1, 0){255}}\put(330,500){\line( 0, -1){
5}}\end{picture}}

\begin{document}
\title{Beyond paired quantum Hall states: parafermions
and incompressible states in the first excited Landau level}
\author{N. Read$^{1,2}$ and E. Rezayi$^{1,3}$} 
\address{$^1$Institute for Theoretical Physics, University of
California,\\ 
Santa Barbara, California 93106-4030}
\address{$^2$ Departments of Physics and Applied Physics, Yale
University,\\ 
P.O. Box 208120, New Haven, Connecticut 06520-8120} 
\address{$^3$ Department of Physics, California State University,\\ 
Los Angeles, California 90032} 
\date{\today} 
\maketitle 
\newcommand{\be}{\begin{equation}}   
\newcommand{\ee}{\end{equation}}
\newcommand{\bea}{\begin{eqnarray}}
\newcommand{\eea}{\end{eqnarray}}
\newcommand{\non}{\nonumber}
\begin{abstract} 
The Pfaffian quantum Hall states, which can be viewed as involving 
pairing either of spin-polarized electrons or of composite fermions, are
generalized by finding the exact ground states of certain Hamiltonians
with $k+1$-body interactions, for all integers $k\geq 1$. The
remarkably simple wavefunctions of these states involve clusters 
of $k$ particles, and are related to correlators of parafermion currents 
in two-dimensional conformal field theory. The $k=2$ case is the Pfaffian. 
For $k\geq 2$, the quasiparticle excitations of these systems are expected to
possess nonabelian statistics, like those of the Pfaffian. For $k=3$, 
these ground states have large overlaps with the ground states of the
(2-body) Coulomb-interaction Hamiltonian for electrons in the first
excited Landau level at total filling factors $\nu=2+3/5$, $2+2/5$. 
\end{abstract} 
\pacs{PACS: 73.40.Hm} 

\narrowtext

%\tableofcontents

%%%%%%%%%%%%%%%%%%%%%%%%%%%%%%%%%%%%%%%%%%%%%%%%%%%%%
\section{Introduction}
\label{introduction}

Trial wavefunctions for fluid states with all the particles in a single Landau
level (LL) have played a paradigmatic role in the fractional quantum Hall 
effect (FQHE) \cite{book} since the work of Laughlin \cite{laugh}. 
The Laughlin states were extended to other filling factors using the hierarchy 
approach \cite{hald,halp84}, which can also be used to generate trial 
wavefunctions, though these are not as elegant or unique as Laughlin's. 
A different motivation for extensions of the Laughlin wavefunctions to yield 
other filling factors is the composite fermion approach \cite{jain}, 
which generates wavefunctions that are simple, but again not quite unique 
for each filling factor. In some cases, trial wavefunctions are also 
energy eigenstates of lowest energy for some Hamiltonian. In the FQHE,
these Hamiltonians have usually been found to annihilate the appropriate trial
states, and whether this is so can be established without much difficulty.
Some early examples are in Refs.\ \cite{hald,haldrez}.

However, these distinct trial wavefunctions at a given filling factor do
not necessarily correspond to distinct phases of matter. Phases of matter
should be characterized by their ground state quantum numbers (including
filling factor or Hall conductance), types of long-range order, ground state
degeneracy (if any), and properties of excitations such as whether or not there
is an energy gap, and the quantum numbers of the {\em elementary} excitations 
that serve as building blocks for all others. In brief, these characteristics
involve {\em universal} aspects of the {\em asymptotic} low-energy,
long-distance physics, and not details of microscopic wavefunctions, either
trial ones or exact energy eigenstates of a realistic Hamiltonian. One usually 
tries to argue that the model Hamiltonian for which the trial ground and 
some excited states are exact eigenstates yields properties that are 
characteristic of a phase. If it is true that the energy spectrum has a gap, 
and hence that the fluids in question are incompressible, then the universal 
properties should be insensitive to perturbations of the Hamiltonian, and 
the properties found {}from the trial states do typify a phase. However, the
existence of a gap is hard to establish for the model Hamiltonians, since it
involves eigenstates of nonzero energy, which are not known exactly. For the
Laughlin state and the pseudopotential Hamiltonian of Haldane \cite{hald}, the
arguments are fairly convincing. (We note that there are also compressible 
liquids \cite{hlr,rr1}, which we believe also to be distinct phases, but for 
which there are so far no Hamiltonians for which exact eigenstates are known. 
The same comment applies to crystalline phases.)  

In the lowest (or ${\cal N}=0$) Landau level (LLL) in single-layer systems, the 
picture that has emerged is that the incompressible fluids are in the phases 
typified by the Laughlin and hierarchy states, or by the composite fermion 
approach, these two approaches yielding the same phase for each filling factor 
with an odd denominator \cite{read90,blokwen}. (There are complications 
involving spin \cite{sondhi}, which we will not discuss here.) In higher LLs, 
the pseudopotentials \cite{hald}, that characterize the interactions within 
the highest partially-filled LL, are different, and the ground states must be 
reconsidered (the relative importance of the Zeeman energy may also change, 
but will not be considered in this paper). For example, the quantized Hall 
plateau (i.e. incompressible fluid) observed at $\nu=5/2=2+1/2$ \cite{willett}, 
where the LLL is filled with electrons of both spin components, and the $1/2$ 
is the filling of the first excited (or ${\cal N}=1$) LL, does not correspond 
to a hierarchy state; no corresponding plateau is seen at $\nu=1/2$ in the LLL. 
A trial state in which the spins of the electrons in the ${\cal N}=1$ LL were 
unpolarized was proposed \cite{haldrez}, however, recent work suggests that the 
true ground state may be polarized \cite{morf,rezhal}. Moreover, even for 
simple fractions such as $2+1/3$, the ground state for the Coulomb-interaction 
values of the pseudopotentials in the ${\cal N}=1$ LL is on the borderline 
between compressible and incompressible states, and the Laughlin state does 
not have such a good overlap with the exact ground state, even when the 
$V_1$ pseudopotential has been increased so as to enter the incompressible 
region \cite{hald}.

It is clear that the study of other trial wavefunctions, that are hopefully
representative of distinct phases {}from the hierarchy, could still be of 
further use in understanding incompressible fluids in higher LLs. One class of 
these is the paired states, in which one attempts to form Laughlin states of 
pairs of electrons, in some sense \cite{halp83,haldrez,mr,gww}. In particular, 
the so-called Pfaffian state was introduced and related to correlation 
functions in two-dimensional conformal field theory (CFT) by Moore and Read 
\cite{mr}. This state occurs for filling of the topmost LL of $\nu=1/q$, for 
$q$ even for fermions, the physical case. Using this correspondence, they 
argued that this incompressible fluid phase has fractionally-charged
excitations of charge $1/2q$ in electron units, and that these possess
nonabelian statistics, a generalization of fractional statistics that we will
discuss later. Also, there are neutral fermion excitations; the fluid ground
state can be viewed as a Bardeen-Cooper-Schrieffer paired state of these
neutral or composite fermions, and the fermion excitations are created by
breaking pairs. The Pfaffian ground state is an exact, zero-energy eigenstate
of a certain Hamiltonian containing a three-body interaction \cite{gww,rr}. 
The claims about the statistics were reinforced by later work which exhibited 
the gapless fermion excitations at the edge \cite{wen,milr}, and the
degeneracy of the quasihole states, for the three-body Hamiltonian 
\cite{nayak,rr}. In the recent work \cite{morf,rezhal}, the true ground state 
for the Coulomb interaction in the ${\cal N}=1$ LL was found to have a sizable 
overlap with the Pfaffian state which can be increased to large values (97 \%) 
as $V_1$ or $V_3$ are varied about their Coulombic values. (Morf \cite{morf}
also found a large overlap with another paired state.) Therefore 
it is possible that the state observed at $\nu=5/2$ is in the phase described 
by the Pfaffian, as first suggested in Ref.\ \cite{gww}.

In this paper we obtain a class of trial wavefunctions by a direct
generalization of ideas that are valid for the Pfaffian state, namely the
zero-energy eigenstates of a $k+1$-body $\delta$-function interaction, where
$k$ is an integer. The existence of the states is established by arguments
using operator product expansions (ope's) in CFT \cite{bpz}, and the explicit 
form of the wavefunctions is obtained and proved to give the ground states. 
The states involve clusters of $k$ particles, generalizing the pairs in the 
Pfaffian state. Further analysis gives predictions about the quasihole states 
and nonabelian statistics, which we partially confirm by solving the special 
Hamiltonians numerically. Then, in Section III we analyse the ground state 
of the Coulomb interaction in the ${\cal N}=1$ LL for $\nu=2+3/5$, and compare 
it with our state at the same filling factor; excellent agreement is found, 
indicating that our functions are serious contenders to describe the phases in 
higher LLs, at least in some cases. Further discussion is given in the 
Conclusion. Some of our results appeared in an earlier unpublished work
\cite{rrunpub}.

%%%%%%%%%%%%%%%%%%%%%%%%%%%%%%%%%%%%%%%%%%%%%%%%%%%%%
\section{Solution of special $k+1$-body interaction Hamiltonians}

In this Section, we show how to solve certain interaction Hamiltonians, in the
sense of finding the zero-energy eigenstates, with the help of operator 
product expansions (ope's) in conformal field theory (CFT) \cite{bpz}. We first
reinterpret the Pfaffian state in this light, then discuss our Hamiltonians. 
Then we analyse the quasihole states, and discuss nonabelian statistics and the
number of sectors of edge states. 

%%%%%%%%%%%%%%%%%%%%%%%%%%%%%%%%%%%%%%%%
\subsection{Notation and Hamiltonians}
\label{Hams}

We will first define some notation for a system of particles in the
lowest Landau level (LLL) on the sphere \cite{hald} (it was used
previously in Ref.\ \cite{rr}).    
The magnetic field is radial and uniform with a total of $N_\phi$ flux
through the surface, and in the lowest Landau level (LLL) each particle 
has orbital angular momentum $N_\phi/2$. The LLL wave functions on a sphere
are usually written (in a certain gauge \cite{hald}) in terms of
``spinor'' (or ``homogeneous'')
coordinates $u_i$ and $v_i$ for each particle $i=1$, \ldots, $N$, with
$u_i=e^{i\phi_i/2}\cos\theta_i/2$, $v_i=e^{-i\phi_i/2}\sin\theta_i/2 $,
where $\theta_i$, $\phi_i$ are the spherical polar coordinates on the
sphere. Since these imply that $u_i$, $v_i$ are not independent complex
numbers, it is often more convenient, and will simplify the writing, to
use a nonredundant parametrization of the sphere by a single complex
variable. This is done by stereographic projection, which gives the
definition $z_i=2Rv_i/u_i$, where $R$ is the radius of the sphere.
Single-particle basis states in the LLL then take the form
$z_i^m/(1+|z_i|^2/4R^2)^{1+N_\phi/2}$, where the $L_z$ angular momentum
quantum number is $L_z=N_\phi/2-m$, so $m\leq N_\phi$. Many-particle
states can thus be written as
\begin{equation}
\Psi=\tilde{\Psi}\prod_i(1+|z_i|^2/4R^2)^{-(1+N_\phi/2)}
\end{equation}
and $\tilde{\Psi}$ must be a polynomial of degree no higher than $N_\phi$ 
in each $z_i$. Therefore, in the following we need specify only
$\tilde{\Psi}$ in order to describe a state. The function $\tilde{\Psi}$
for a ground state on the sphere can also be used to construct a
wavefunction suitable for a corresponding disk-shaped ground state on the
plane, by multiplying by $\exp(-\frac{1}{4}\sum_i |z_i|^2)$.

The Pfaffian state is defined as \cite{mr}
\begin{equation}
\tilde{\Psi}_{\rm Pf}(z_1,\ldots,z_{N})=\hbox{Pf}
\left({1\over z_i-z_j}\right)\prod_{i<j}(z_i-z_j)^q. 
\label{pfaffstate}
\end{equation}                                                  
The Pfaffian 
\be
\hbox{Pf}\,M_{ij}=(2^{N/2}(N/2)!)^{-1}\sum_P\hbox{sgn}\,P
\prod_{r=1}^{N/2}M_{P(2r-1)P(2r)}
\ee
of an antisymmetric $N\times N$ matrix $M$ ($N$ even) is the antisymmetrized 
sum over all pairings $(z_i-z_j)^{-1}$ (the analogous pairing in the 
spin-singlet case appears in the Haldane-Rezayi state \cite{haldrez}). 
The filling factor of the state is $\nu=N/N_\phi\rightarrow 1/q$ as 
$N\rightarrow \infty$ (since $N_\phi=q(N-1)-1$), and $q$ is odd for a boson 
state and even for fermions. 

The Pfaffian state for $q=1$, $q=2$ is the ground state of a three-body
Hamiltonian \cite{gww,rr}. For these cases, the Hamiltonian penalises the
closest approach of any three particles allowed by the statistics. Thus
for $q=1$, where the particles are bosons (note that we reserve the term
``particles'' for the underlying particles, which are either charged
bosons or charged fermions [electrons], and not for composite particles),
the Hamiltonian can be taken to be \cite{gww}  
\begin{equation}
H=V \sum_{i<j<k}\delta^{2}(z_{i}-z_{j})\delta^{2}(z_{i}-z_{k}),
\label{pfaff3bodH}
\end{equation}
where the sum is over distinct triples of particles \cite{note}.
For numerical purposes on the sphere, it is more convenient to work with a
projection operator form of the three-body Hamiltonian, instead of
the $\delta$-functions in (\ref{pfaff3bodH}). The closest
approach of three particles on the sphere corresponds to the state of
maximum possible total angular momentum for the three. If the particles
are bosons, the largest possible total angular momentum is $3N_\phi/2$
(recall that each particle has angular momentum $N_\phi/2$). Then, for
the $q=1$ case, the Hamiltonian may be taken as proportional to the
projection operator onto the (unique) multiplet of maximum angular momentum 
for each triple of bosons:
\begin{equation}
H=\sum_{i<j<k}V P_{ijk}(3N_\phi/2).
\label{pfaff3bodprojH}
\end{equation}  
The same trick works for the three-body interaction of fermions giving
the $q=2$ case; in this case, the maximum total angular momentum of three
particles is $3N_\phi/2-3$. On the plane, the latter Hamiltonian
corresponds to derivatives of delta functions. For these two cases, the
Pfaffian state on the sphere is the unique energy eigenstate of 
zero energy at the stated $N_\phi$ value. (We will refer to such states 
as zero-energy states hereafter.)  For larger $q$, these Hamiltonians
can be generalized, in such a way that the zero-energy states are obtained
{}from those for $q=1$ by multiplying by $\prod(z_i-z_j)^{q-1}$ (it is
assumed that for $q$ odd, we are discussing bosons, and for $q$ even,
fermions). The presence of the latter factor implies that they are all
zero-energy eigenstates of the projection operators for any two particles
onto relative angular momentum $M=0$, $2$, \ldots, $q-3$ ($q$ odd), or
$M=1$, $3$, \ldots, $q-3$ ($q$ even) [or the corresponding total angular
momenta $N_\phi$, $N_\phi-2$, \ldots, $N_\phi-q+3$, ($q$ odd), etc.]. 
The space of states annihilated by such projections is in one-to-one
correspondence with the full space of states of the $q=1$ case, and the
desired three-body projection operator [onto angular momentum
$3N_\phi/2-3(q-1)$] is the unique one that corresponds under this mapping
to that already mentioned for $q=1$. For each $q$, the Hamiltonian can
then be taken to be the sum of the three-body and all of these two-body
projection operators. A very similar approach works for the other
Hamiltonians studied in this paper, so that results for higher $\nu^{-1}$
can be deduced easily {}from those for the minimal $\nu^{-1}$ for each
type of state. These Hamiltonians can also be written in terms of
$\delta$-functions and their derivatives, so as to arrive at a form
suitable for use in geometries other than the sphere.

The goal of this paper can now be stated: we wish to generalise the
solution for the zero-energy states of the three-body Hamiltonian
(\ref{pfaff3bodH}) or (\ref{pfaff3bodprojH}) to $k+1$-body
Hamiltonians of the same closest approach form, for every $k$ (note that
for $k=1$, we can consider the solutions to be the Laughlin ground and
quasihole states \cite{hald}). We will demonstrate the existence of an
elegant solution to this problem by using conformal field theory,
construct the ground state wavefunctions explicitly, and then compare the
states with solutions for the Coulomb interaction in the second LL.

%%%%%%%%%%%%%%%%%%%%%%%%%%%%%%%%%%%%%%%%%%
\subsection{Ope's and the Pfaffian state}
\label{opes}

First we will reconsider the Pfaffian state and show how its property of
being the unique zero-energy state for a certain Hamiltonian, that is of
vanishing as three particles come to the same point, is related to
operator product expansions in the corresponding CFT. We then generalize
this to find the solutions to the simplest examples of $k+1$-body
Hamiltonians, for each $k$. The use of CFT leads to an existence proof for
these wavefunctions, and in principle determines the wavefunctions
uniquely.

The basic idea of Ref.\ \cite{mr} is that the {\em wavefunction} of 
our 2 space-, 1 time-dimensional system can be
related to a {\em correlator} of a certain conformal field theory.
Here we are not interested in the simple Laughlin-Jastrow factors, but 
in the other parts that may produce nonabelian statistics. For the
Pfaffian case, this part is the theory of free massless
Majorana Fermi fields in 2 Euclidean spacetime dimensions \cite{mr}:
\begin{equation}
\tilde{\Psi}_{\rm Pf}=\langle\psi(z_1)\cdots\psi(z_N)\rangle
\prod_{i<j}(z_i-z_j)^q,     
\end{equation}
in which the correlator can be evaluated using Wick's theorem and 
$\langle\psi(z)\psi(w)\rangle$ $=
-\langle\psi(w)\psi(z)\rangle$ $= (z-w)^{-1}$. The Fermi 
fields can also be characterized by the operator product expansions (ope's)
which hold inside correlators \cite{bpz}
\begin{eqnarray}
\psi(z)\psi(w)&\sim&(z-w)^{-1}[I+2(z-w)^2T(w)+\ldots],
\label{psipsiope}\\
\psi(z)I(w)&\sim&\psi(w)+(z-w)\partial\psi(w)+\ldots,
\label{psiIope}
\end{eqnarray}
as $z\rightarrow w$, where $I$ is the identity (which is a constant), and 
$T(z)=-\frac{1}{2}\psi\partial\psi$ is the stress tensor. Throughout, 
$\partial\phi(z)$ will mean $\partial\phi(z)/\partial z$ and dots $\dots$ will 
denote higher order, less-singular terms. While the correlator of Majorana 
fermions can be found by Wick's theorem, the use of ope's is more general 
and can be applied in a vast number of situations where there is no simple 
Wick's theorem.  

We now show how the ope for the fermions guarantees that the state is a
zero-energy eigenstate for the three-body interaction of Greiter, Wen and
Wilczek \cite{gww}. This can be seen directly \cite{gww} {}from the explicit
wavefunction (\ref{pfaffstate}), however, the ope's provide a {\em general} 
argument that can be used {\em even when the wavefunctions are unknown}, if 
the function is a correlator of fields with known ope's. Suppose that in the 
correlator (without loss of generality) first $z_2\rightarrow z_1$, then
$z_3\rightarrow z_1$, and take the most singular term of each product. 
The first limit (using Eq.\ (\ref{psipsiope})) gives $(z_1-z_2)^{-1}I$,
and the second (using Eq.\ (\ref{psiIope})) then gives 
$(z_2-z_1)^{-1}\psi(z_1)$. Multiplying by the
Laughlin-Jastrow factor $\tilde{\Psi}_{\rm LJ}=\prod_{i<j}(z_i-z_j)$ we 
find that $\tilde{\Psi}_{\rm Pf}$ for $q=1$ vanishes as $(z_3-z_1)(z_3-z_2)$, 
i.e.\ quadratically. Hence for the short-range interaction among bosons
in Eq.\ (\ref{pfaff3bodH}) we obtain zero energy \cite{gww}. Furthermore
it is the densest such state \cite{gww,milr,rr}. The same argument applies 
for $q>1$ by construction.

%%%%%%%%%%%%%%%%%%%%%%%%%%%%%%%%%%%%%%%%%%%%%%%%%%
\subsection{Parafermion states}
\label{parafermion}

We next generalize the ideas to solve Hamiltonians with $k+1$-body
interactions. The simplest such Hamiltonian, which will give the highest
density such state for each $k$, which is always a state for bosons, is
a $\delta$-function interaction between the $k+1$ particles:
\widetext
\top{-2.8cm}
\be
H=V\!\!\!\sum_{i_1<i_2<\cdots,i_{k+1}}\delta^{2}(z_{i_1}-z_{i_2})
\delta^{2}(z_{i_2}-z_{i_3})\cdots\delta^2(z_{i_{k}}-z_{i_{k+1}}).
\ee
As we will see, the wavefunctions involve dividing the particles into
clusters of $k$ each. To generalize the approach just described for the
Pfaffian, we will consider a natural generalization of the ope algebra of
Majorana fermions, which is the algebra of ${\bf Z}_k$ parafermions
\cite{zamo}. This consists of a set of 
fields $\psi_\ell(z)$, $\ell=1$, \ldots, $k-1$ with ope's
($\psi_\ell^\dagger(z)=\psi_{k-\ell}(z)$)
\bea
\psi_\ell(z)\psi_{\ell'}(z')&\sim&
c_{\ell,\ell'}(z-z')^{-(\Delta_\ell+\Delta_{\ell'}-\Delta_{\ell+\ell'})}
%\non\\&&\mbox{}\times
[\psi_{\ell+\ell'}(z')+\ldots]\quad(\ell+\ell'<k),
\label{paraope}\\
\psi_\ell(z)\psi_{\ell'}^\dagger(z')&\sim&c_{\ell,k-\ell'}
                           (z-z')^{-(\Delta_\ell+\Delta_{\ell'}
        -\Delta_{\ell-\ell'})}
%\non\\&&\mbox{}\times
[\psi_{\ell-\ell'}(z')+\ldots]\quad(\ell'<\ell),\\
\psi_\ell(z)\psi_{\ell}^\dagger(z')&\sim&(z-z')^{-2\Delta_\ell}
%\non\\&&\mbox{}\times
[I+\frac{2\Delta_\ell}{c}(z-z')^2T(z')+\ldots],\\
T(z)\psi_\ell(z')&\sim&\frac{\Delta_\ell}{(z-z')^2}\psi_\ell(z')
                     +\frac{1}{z-z'}\partial\psi_\ell(z')+\ldots.
\eea
The consistency of the algebra puts conditions on the conformal weights 
$\Delta_\ell$ of the $\psi_\ell$'s. The simplest solution to the conditions 
is $\Delta_\ell=\ell(k-\ell)/k$, and with this choice the values of 
the central charge $c=2(k-1)/(k+2)$ and the numerical
coefficients $c_{\ell,\ell'}$ are then uniquely determined by consistency. 
With these choices, we obtain the algebra usually known as ${\bf Z}_k$
parafermions.
(Other solutions for the $\Delta_\ell$ are listed in Ref.\ \cite{zamo}, and
some are studied further. In these cases the other parameters are not uniquely
determined. These other algebras will not be needed here, but may be relevant 
in other FQHE problems. The procedure below will generate wavefunctions in all
cases, but these may not always be solutions to simple Hamiltonians.)
For $k=2$, $\psi_1=\psi$ is a Majorana fermion, and the
ope's are the same as Eqs.\ (\ref{psipsiope},\ref{psiIope}) above.

Consider the function
\begin{equation}
\tilde{\Psi}_{\rm para}^{(m)}(z_1,\ldots,z_N)=\langle\psi_1(z_1)\cdots
                       \psi_1(z_N)\rangle\tilde{\Psi}_{\rm LJ}^{M+2/k}
\label{psipara}
\end{equation}
where $N$ is divisible by $k$ and $M\geq0$ is an integer. The 
$\tilde{\Psi}_{\rm LJ}^{M+2/k}$ factor
renders (\ref{psipara}) nonsingular $\sim(z_i-z_j)^M$ as any $z_i\rightarrow 
z_j$, according to Eq.\ (\ref{paraope}) with $\Delta_\ell+\Delta_{\ell'}
-\Delta_{\ell+\ell'}=2\ell\ell'/k$, which gives $2/k$ for $\ell=\ell'=1$. 
$\tilde{\Psi}_{\rm para}$ is totally symmetric for $M$ even, antisymmetric
for $M$ odd, so describes bosons or fermions, respectively. As any 
$z_i\rightarrow\infty$, the correlator $\sim z_i^{-2\Delta_1}$, 
because the separations among the other $N-1$ $\psi_1$'s are then
relatively small, and the ope's imply that they ``fuse'' to form a
single $\psi_1^\dagger$. Hence $\tilde{\Psi}_{\rm para}$ is a 
polynomial of degree $N_\phi=\nu^{-1}N-(M+2)$, where $\nu=k/(Mk+2)$ 
is the filling factor. 

Taking now the $M=0$ case, we can show that the function vanishes
quadratically as any $k+1$ particles come to the same point, say
$z_2$, \ldots, $z_{k+1}$ approach $z_1$ one by one, by a similar argument
based on the ope's as for the Pfaffian. As the first $k-1$ particles
approach $z_1$, the most singular (or most slowly vanishing) terms must
give the same result as the $N=k$ case of $\tilde{\Psi}_{\rm para}$, which
is a constant \cite{zamo}; one additional particle gives
$\psi_1(z_{k+1})I\sim\psi_1(z_1)$, and the LJ factors give
$(z_{k+1}-z_1)^2$. Hence like the Pfaffian state with $q=M+1=1$, this
state is a zero-energy state for the $k+1$-body $\delta$-function
interaction of bosons, and this fact again extends to $M>0$.
This shows that the desired wavefunctions exist, provided the CFT
correlators do; but the parafermion theories are well-studied and can be
constructed {}from other well-understood theories \cite{zamo,gepqiu}, so
correlators with the properties specified by the ope's do exist. Notice that,
had we chosen to use a different parafermion operator $\psi_\ell$ ($\ell\leq
k/2$) in place of all the $\psi_1$'s, we would have obtained a polynomial 
that vanishes more rapidly when $k+1$ $z_i$'s coincide, and a lower filling 
factor.

Explicit functions can be obtained using the idea of grouping particles into
clusters of $k$ \cite{halp83}. The following procedure gives a symmetric 
$\tilde{\Psi}^{(0)}_{\rm para}$ of the correct degree which is the Laughlin 
state for $k=1$, and the Pfaffian state for $k=2$: Divide the particles
into clusters of $k$. For each pair of distinct groups, say $z_1,\ldots,z_k$ 
and $z_{k+1},\ldots,z_{2k}$, we define factors $\chi$ by (for $k\geq2$)
\be
\chi(z_1,\ldots,z_k;z_{k+1},\ldots,z_{2k})=
%&&\non\\
%&&\quad
(z_1-z_{k+1})(z_1-z_{k+2})(z_2-z_{k+2})(z_2-z_{k+3})
%\non\\
%&&\quad\quad\mbox{}
\cdots(z_k-z_{2k})(z_k-z_{k+1}), 
\ee
that is, each member of a cluster is connected with only two
members of the other cluster, by a factor $z_i-z_j$. For $k=1$, we would have
$\chi=(z_1-z_2)^2$. Now we multiply these for all distinct pairs of clusters, 
and symmetrize the whole expression, to obtain
\be
\tilde{\Psi}_{\rm para}^{(0)}=
{\sum_{P\in S_N}}^\prime\prod_{0\leq r<s<N/k}
\chi(z_{P(kr+1)},\ldots,z_{P(k(r+1))};z_{P(ks+1)},\ldots,z_{P(k(s+1))})
\label{parawfn}
\ee
\bottom{-2.7cm}
\narrowtext
\noindent
where $S_N$ is the permutation group on $N$ objects, and the prime on the
sum denotes that the summation can be restricted to permutations obeying
$P(1)<P(k)<\cdots<P(N-k+1)$, which eliminates redundant permutations of
the clusters; otherwise each term would appear $(N/k)!$ times (in the
expression for the Pfaffian above, we instead summed
over all permutations but divided by the number of times each distinct
product appeared). Hence this function is totally
symmetric, even with this restriction. To obtain larger $M>0$, we must
multiply by $\tilde{\Psi}_{\rm LJ}^M$. The degree of the function in each
$z_i$ is then $N_\phi=M(N-1)+2(N/k-1)=\nu^{-1}N-(M+2)$, as for the
parafermion correlator times LJ factor above. 
%Our construction is a 
%natural generalization of the Laughlin state where there is only a single 
%particle in each group and $\chi(z_1,z_2)=(z_1-z_2)^2$.

Next we show that our function (for $M=0$) vanishes quadratically whenever
any $k+1$ particles come to the same point. It is sufficient to consider
each term in the sum over permutations separately. Clearly, we must then
consider many possibilities, in which the $k+1$ particles are distributed
among different clusters, ranging {}from all but one in the same cluster, to
each in distinct clusters. We will organise our proof in the following way:
first we show that each term vanishes if $k$ particles come to the
same point, except in the case of all $k$ in the same cluster. Then we
use the evident fact that for $k$ particles in one cluster, one in another,
the term vanishes (this follows directly {}from the definition of $\chi$
itself). 

We must consider $k$ distinct $z_i$'s, chosen in any way {}from up to $k$
distinct clusters. We may label these by the corresponding position in a
cluster ({}from 1 to $k$), and the label (like $r$, $s$) for the cluster, that
appear inside the permutation operator $P$. These determine when the
product of $\chi$'s vanishes. Put another way, it is sufficient to
consider the term where $P$ is the identity. Now consider the values for
the position within a cluster as arranged on a circle (or clock face),
with the numbers increasing clockwise until $k$ is reached, which is
followed by $1$, so that 1 is adjacent to 2 and $k$, 2 is adjacent to 1
and 3, and so on. The $\chi$'s containing any two selected $z_i$'s vanish
only in two cases: (i) if members of distinct clusters occur at the same
place on the clock; (ii) if members of distinct clusters occur at adjacent
positions, but only if the position of the later cluster is arrived at by
moving clockwise {}from the earlier by one step (see the definition of
$\chi$). We can then describe the structure of the term containing any
set of $k$ $z_i$'s by labelling positions on the clock face with the
numbers of the clusters in which the $z_i$'s appear. Members of the same
cluster must be at different clock positions because they are distinct
$z_i$'s. Clearly, if a term is to have any possibility of not vanishing,
we must choose all the clock positions to be distinct. We then have a
single number of a cluster assigned to each clock position. We will now show
that it is not possible to arrange these in such a way that the term does not
vanish, except by choosing all the $z_i$'s {}from a single cluster. We can
view the clock as divided into regions (possibly consisting of positions
that are not all adjacent) that have been labelled (or ``colored'') with a
single cluster number. To avoid a vanishing factor, the numbers of the
clusters must decrease as one moves clockwise {}from one region to another.
But the clock is not simply connected, and so this cannot be done all the
way around the clock, unless there is only a single region (or cluster)
involved. In the latter case, consideration of any one other $z_i$ shows
that the term vanishes, as mentioned before. Therefore, we have shown that
the function vanishes at least linearly, but since it is totally symmetric
after summing over permutations, it will actually vanish quadratically as
any $k+1$ particles come to the same point.     

We have shown that this vanishing property follows {}from the 
parafermion ope's, which should determine the function, however the vanishing
property alone may be only a necessary and not a sufficient condition to 
determine that the function is the parafermion correlator times 
$\tilde{\Psi}_{\rm LJ}^M$. Nonetheless, since we have demonstrated it for 
an explicit function, and since this function seems to be unique, on the basis 
at least of numerical verification for $k=3$, $4$ (discussed further below), 
we conjecture that the parafermion correlators for all $k$ are given
by this construction (by dividing $\tilde{\Psi}^{(0)}_{\rm para}$ by
$\tilde{\Psi}_{\rm LJ}^{2/k}$). It should be possible to prove or disprove this
statement by studying the properties of our function as various combinations of
$z_i$'s coincide, and comparing these with the ope's. This was done for various
paired states in Ref.\ \cite{wenwu}, but we will not consider it further.

%%%%%%%%%%%%%%%%%%%%%%%%%%%%%%%%%%%%%%%
\subsection{Quasihole states and nonabelian statistics}
\label{qholes}

In this Subsection, we argue that states containing quasiholes (carrying
charge $1/(Mk+2)$ each) for the parafermion states with 
$k>2$ can be constructed in analogy with the Pfaffian case, and that 
nonabelian statistics are expected in all cases. The arguments go as follows.  
We will again begin with the Pfaffian state.

The two-quasihole wavefunction proposed in Ref.\ \cite{mr} was 
\begin{equation}
\tilde{\Psi}_{\rm Pf+2\,qh}(z_1,\ldots;w_1,w_2)=
\hbox{Pf}\left({f(z_i,z_j;w_1,w_2)\over
z_i-z_j}\right)\tilde{\Psi}_{\rm LJ}^q 
\label{2qhole}
\end{equation}
where $f(z_i,z_j;w_1,w_2)=(z_i-w_1)(z_j-w_2)+(z_i-w_2)(z_j-w_1)$. 
It can be interpreted as the insertion of two spin fields $\sigma(w)$:
\bea
\tilde{\Psi}_{\rm Pf+2\,qh}&\propto&
\langle\psi(z_1)\cdots\psi(z_N)\sigma(w_1)\sigma(w_2)\rangle\non\\
 &&  \mbox{}\times\tilde{\Psi}_{\rm
LJ}^q\prod_i(z_i-w_1)^{1/2}(z_i-w_2)^{1/2}.
\eea
The spin fields induce square-root branch singularities in the fermi
fields, described by the ope's \cite{bpz}
\begin{equation}
\psi(z)\sigma(w)\sim(z-w)^{-1/2}\sigma(w)+\ldots.
\label{spinpsiope}
\end{equation}
The branch singularities are cancelled by the explicit square roots, to
ensure that the wavefunctions are single-valued. This fixes the charge of
each quasihole to be $1/2q$. We note \cite{mr} that quasiholes can be
created only in pairs. 

In Ref.\ \cite{mr} it was proposed to extend this
by defining wavefunctions for $2n$ quasiholes by inserting $2n$ spin
fields. Since the ope's, by definition, describe short distance properties
of correlators independently of what other fields are present, our
argument above implies that {\em the quasihole wavefunctions so obtained
will all be zero-energy eigenstates of the appropriate $H_3$} (this can
again be seen explicitly in the $n=1$ case above). This is analogous to
the Laughlin states, where for a suitable pseudopotential Hamiltonian the
ground state and states with quasiholes added are zero-energy states
\cite{hald}. However, for the present case, there is not just a unique
state for each set of quasihole positions. Instead there are 
$2^{n-1}$ linearly-independent functions (conformal blocks) of the $z_i$'s
for $2n$ spin fields. This number follows {}from the ope of 
the spin fields \cite{bpz}:
\begin{equation}
\sigma(z)\sigma(w)\sim(z-w)^{-1/8}I+{\rm const}(z-w)^{3/8}\psi(w)+\ldots.
\label{spinope}
\end{equation}
Here the higher-order terms fall into two families since the behaviour of
each term as $z\rightarrow w$ differs {}from that of one of the first two terms 
by an integer power of $(z-w)$. Then in a correlator containing
$\sigma(w_1)\sigma(w_2)\cdots$, as $w_1\rightarrow
w_2$, $w_3\rightarrow w_4$, etc we can choose a member of either set of 
terms for each pair of $\sigma$'s, which would give $2^n$ terms, except that 
the total number of $\psi$'s in the correlator must be even, so we get 
$2^{n-1}$ blocks, for $N$ (the number of $\psi$ insertions) odd or even. 
In principle, the ope's can be used to fix the actual form of the
correlators for $2n\geq 4$. In practice, it was more convenient to find
all the zero-energy states by explicit construction, and it was shown that
for fixed positions of the quasiholes, the number of states is $2^{n-1}$
\cite{nayak,rr}. Thus the results for the Pfaffian
are in complete agreement with these predictions based on CFT ope's.

These degenerate spaces of quasihole states are the basis for the
nonabelian statistics properties of the quasiholes (and similar properties
are expected for quasielectrons, or for combinations of quasiholes and
quasielectrons). As the locations $w_i$ of the spin fields are exchanged
{\em by analytic continuation}, the conformal blocks are mapped to linear
combinations of each other (monodromy) \cite{bpz}. The conjecture of
Ref.\ \cite{mr} was that when the quasiholes are exchanged {\em
adiabatically} as in Ref.\ \cite{asw}, these functions exhibit nonabelian
statistics, meaning that the effect on members of the space is a linear
transformation, which can be described by a $2^{n-1}\times2^{n-1}$ matrix
by choosing a basis, and that this is the same as the matrix obtained by
analytic continuation. This replaces the usual Berry phase 
that describes ordinary (abelian) fractional statistics; matrices representing
distinct exchanges do not usually commute, hence the term ``nonabelian''. The 
existence of many blocks is thus a necessary but not sufficient condition
for nonabelian statistics (the adiabatic exchange has not yet been
calculated explicitly) \cite{note2}.

We now discuss the extension of these results to the parafermion states
for $k>2$. In place of the spin fields for the Majorana fermion, the
parafermion system has ``fields'' (chiral vertex operators) $\Phi^l_m$ 
\cite{zamo,gepqiu}, where $l=0$, $1$, \ldots, $k$, while $m$ is a
periodic variable with period $2k$, so $m=0$, $1$, \ldots, $2k-1$, and
further $l-m=0$ (mod 2). In the ope's of $\Phi^l_m$, the $m$'s
add mod $2k$, however, there are also identifications which imply that
$\Phi^{k-l}_{k+m}=\Phi^l_m$ are the same operator, so
the values of $m$ can be restricted so as to get each of the
$k(k+1)/2$ distinct fields once;
a convenient way to do this is by restricting $-l<m\leq l$. In this
notation, $\Phi^0_0=\Phi^k_k=I$ and
$\Phi^k_{2\ell-k}=\Phi^0_{2\ell}=\psi_{\ell}$.
The special cases $\sigma_l=\Phi^l_l$, $l=1$, \dots, $k$, ($\sigma_k=I$),  
are called primary fields for the parafermion algebra. $\sigma_1$ 
in a sense generates the whole set by repeated operator products, and is a
natural analogue for $\sigma$, to which it reduces for $k=2$. We propose
to insert a number $nk$ of $\sigma_1$'s into the correlator of parafermion
currents to obtain the basic quasihole states; these are zero-energy
states for our special Hamiltonians by the preceding argument. We use
$\sigma_1$ because this leads to the minimal charge for the quasiholes.
Thus our proposal for quasihole states is to use  
\bea
\lefteqn{\tilde{\Psi}^{(M)}_{\rm
para+qh}(z_1,\ldots;w_1,\ldots,w_{nk})=}\non\\
&&\quad\langle\psi_1(z_1)\cdots\psi_1(z_N)\sigma_1(w_1)\cdots
    \sigma_1(w_{nk})\rangle\non\\
&&\quad\quad\mbox{}\times\tilde{\Psi}_{\rm
LJ}^{M+2/k}\prod_{i=1}^N\prod_{p=1}^{nk}
(z_i-w_p)^{1/k}.
\eea
In these states, $N_\phi=\nu^{-1}N-(M+2)+n$.
Once again, this will in general define a whole set of states (conformal
blocks), not a unique state. In the last factor, the exponent $1/k$ is
chosen to cancel the branch singularity as any $\psi_1$ approaches any
$\sigma_1$, which is determined by the ope,
\be
\psi_1(z)\Phi^1_1(0)\sim z^{-1/k}\Phi^1_3(0)+\ldots
\ee
(see Refs.\ \cite{zamo,gepqiu}). This is the weakest singularity for any 
of the $\Phi^l_m$, so gives the smallest charge for a quasihole. The
charge on the quasiholes is determined entirely by the Laughlin-like part
of the wavefunction, not the conformal block of the parafermions, so the
usual plasma argument shows that the charge (number of particles missing
{}from the vicinity of $w_p$) is $\nu/k=1/(Mk+2)$ (use of the other spin
fields $\sigma_l$ in general give charges $\nu l/k$, and by including
additional Laughlin quasihole factors we can add integer multiples of
$\nu$ to these charges). $Mk+2$ is the denominator
of $\nu=k/(Mk+2)$, however $k$ and $Mk+2$ have a common factor if and
only if $k$ is even, and this factor is just 2 (e.g. for the Pfaffian,
$k=2$). Therefore, when $k$ is even, the
charge is fractionalized compared with the usual value in a Laughlin or
hierarchy state for a spin-polarized single component system at the same
filling factor, which is always $1/q$ for filling factor $p/q$, where $p$,
$q$ have no common factors. The next nontrivial example is $k=4$, where
with $M=1$ for fermions we obtain a $\nu=2/3$ state, or $\nu=1/3$ by
particle-hole inversion, with charge $\pm 1/6$ excitations. 

The explicit function for $n=1$, that is $k$ quasiholes, is the same as
$\tilde{\Psi}_{\rm para}^{(0)}$, except that a factor of the form
\be
\Gamma(z_1,\ldots,z_k;w_1,\ldots,w_k)=\prod_i(z_i-w_i)
\ee
for each cluster of $k$ particles is inserted inside the sum on permutations 
$P$ in Eq.\ (\ref{parawfn}); the permutations act only on the $z_i$'s, not the
$w_p$. This generalizes the function $f$ above for the 
Pfaffian, except that once again the symmetrization is now done all at once by 
the sum over $P$, which can still be restricted as before. The resulting
function is also symmetric in the $w_p$'s.  

For more quasiholes, we do not have the explicit functions in general, 
but we can count the number of zero-energy states for fixed positions of
the quasiholes in the above construction, using the CFT, in a similar way 
as for the Pfaffian state above; we will consider
$k=3$ explicitly. We require the ope's of the fields $\Phi^l_m$, which are
given in Ref.\ \cite{gepqiu} for all $k$, and will not be written down
explicitly here. Using the ope's one finds by repeatedly taking ope's with
$\sigma_1$, that the number of conformal blocks in the parafermion theory
for $k=3$ for $3n$ quasiholes is a Fibonacci number, $F_{3n-2}$, where we
define $F_1=1$, $F_2=2$, $F_3=3$, $F_4=5$, and $F_m=F_{m-1}+F_{m-2}$ in
general \cite{schoutens}. Thus, the number of linearly-independent 
zero-energy states for our Hamiltonian is also (at least) $F_{3n-2}$ for fixed 
positions of $3n$ quasiholes, in the $k=3$ case, provided $N$ is sufficiently 
large. For large $n$, this number approaches $\sim (2+\sqrt{5})^n$. As $k$
increases, the expressions for the $n$ dependence for the parafermion
cases will become progressively more complex than that for the Pfaffian
($k=2$) case, which was $2^{n-1}$. 

The results for the quasiholes can be compared with numerical results for
the four-body Hamiltonian. The results for the total number of zero-energy 
states (compare Ref.\ \cite{rr} for results on the Pfaffian state),
were calculated for $N=6$, $9$, and $12$ fermions at $\nu=3/5$, but should
be independent of $M$. First, the ground state at the stated $N_\phi$ is
unique, and for small sizes was verified to be given by the
explicit polynomial above. For one flux added ($n=1$), that is $3$ quasiholes, 
the number of states was $${{N/3+3}\choose 3},$$ as one would expect for $3$
bosons in $N/3+1$ orbitals, or {}from the explicit wavefunctions above. For
$n=2$ flux added, that is $6$ quasiholes, the numbers of states were in
agreement with the formula 
\be
{{N/3+6}\choose 6 }+3 {{N/3+5}\choose 6}+{{N/3+4}\choose 6},
\ee
which is similar in form to those found for the Pfaffian state \cite{rr}.
Assuming this works for all larger $N$, and dividing by the first term
which is the value of the positional degeneracy that would be expected if
the quasiholes obeyed abelian statistics \cite{rr}, we obtain the ratio
$5$ in the thermodynamic limit, which agrees with the number expected {}from
conformal field theory. Similarly, for $9$ quasiholes ($n=3$), the results 
agree with the formula
\be
{{N/3+9}\choose 9}+10{{N/3+8}\choose 9}+10{{N/3+7}\choose 9}
\ee
and so we expect the ratio to be 21 in the thermodynamic limit, as 
expected for fixed positions of the quasiholes. For more quasiholes,
because of size limitations we have not been able to obtain any 
such formulas. Thus we find a satisfying agreement
with our prediction, which tends to confirm that all the zero-energy
quasihole states are obtained by inserting $\sigma_1$'s in the parafermion
correlator.

We have not obtained results for the 4-body Hamiltonian ground states on
the torus, or the edge excitations (compare \cite{milr,rr}), either
analytically or numerically. However, we expect that these calculations
would lead in general to the conclusion that the number of sectors of 
edge states, or ground states on the torus, is $(k+1)(Mk+2)/2$ (notice 
that this integer is divisible by the denominator of the filling factor 
in all cases, as required by Ref.\ \cite{hald85}). This result is based 
on a natural structure for the CFT, including the U($1$) charge sector as well 
as the parafermion sector. Our analysis of this theory, into which we will not 
go in detail here, also indicates that while for $k=2$, there are neutral 
fermion excitations, both at non-zero energy in the bulk and as gapless 
excitations at the edge \cite{mr,gww,milr}, for $k>2$ the analogous 
parafermion $\psi_1$ excitations carry charge $2\nu/k$ plus multiples of $\nu$.
The excitation containing a $\psi_1$ and charge 1 is identified with the 
physical hole, as usual, and in the CFT interpretation that applies 
(for example) to the CFT of edge excitations, it generates the chiral algebra, 
as in the Pfaffian case \cite{mr,milr}. There are, however, neutral parafermion 
excitations for $k$ even, that contain $\psi_{k/2}$. These can be viewed as 
being made {}from $k/2$ particles and $(Mk+2)/2$ flux. These are simply 
fermions for $k/2$ odd, bosons for $k/2$ even, and are like the usual composite 
particles. Also, for all $k$, there are nontrivial neutral excitations 
originating {}from the spin fields, $\Phi^l_0=\Phi^{k-l}_k$ in the earlier 
notation. For $k=3$, the latter, $\Phi^2_0$, is the only field other than the 
parafermions $\psi_\ell$ and the spin fields $\sigma_l$. These have no 
analogues either in the hierarchy or the Pfaffian ($k=2$) states, though they 
do in the generalized hierarchy states, such as the 331 state \cite{milr}. We 
further note some isomorphisms of the algebraic structures. For $M=0$, the 
full chiral algebra, including the U(1) charge sector, is the level $k$ 
SU($2$) Kac-Moody current algebra, with representations (sectors) labelled by 
``spin'' $j=l/2=0$, $1/2$, \ldots, $k/2$ (this spin is of course not the
physical spin, which is always polarized). For $M=1$, the chiral algebra is the 
so-called $N=2$ superconformal algebra in the antiperiodic sector, and the 
representations for each $k$ make up the known discrete series $k=1$, $2$, 
\ldots, for this algebra \cite{bfk}. In these cases, we have simply recovered 
known constructions of these algebras and representations {}from the 
parafermions \cite{zamo,zamo2}. The special cases $k=1$ (the Laughlin state) 
and $k=2$ (the Pfaffian state) of these were mentioned earlier \cite{mr,milr}, 
and the SU($2$) $k=2$, $M=0$ case was used in a recent paper \cite{frad}.

%%%%%%%%%%%%%%%%%%%%%%%%%%%%%%%%%%%%%%%%%%%%%%%%%%
\section{Coulomb interaction in the first excited Landau level}
\label{numerical}

We next turn to finite-size calculations.  We have numerically constructed 
the wavefunctions of parafermion states for $\nu=3/5$ ($k=3$, $M=1$), and
$\nu=2/3$ ($k=4$, $M=1$) for small sizes on the sphere and have confirmed
that the $k+1$-body Hamiltonians possess unique zero-energy ground states
at the given flux $N_\phi$. For $k=3$, we have also obtained the
excitation spectrum, both for the model $k+1$-body interaction and the
Coulomb potential in the ${\cal N}=1$ Landau level, and have studied the 
overlaps of the Coulomb ground state with our state as well as with the usual
hierarchy states as the pseudopotential $V_1$ is varied about the
its (${\cal N}=1$) Coulombic value.  Below we describe these results.

\begin{figure}
\inseps{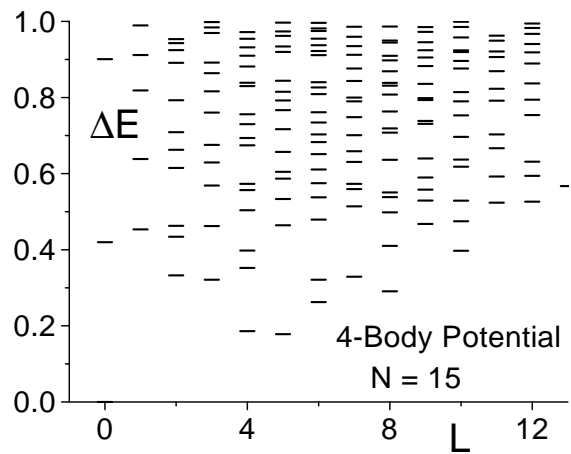}{0.35}
\caption{The low-lying spectrum for the four-body Hamiltonian
for $N=15$ electrons. The ground state is at $L=0$, $\Delta E=0$.}
\label{fig:specM}
\end{figure}

Fig.~\ref{fig:specM} shows the low-lying spectrum for $N=15$ and $\nu=3/5$ for 
the four-body Hamiltonian. The ground state is at $L=0$,
zero energy. The low-lying spectrum bears some resemblance to the
``hanging chain'' shape seen in paired systems.  

\begin{figure}
\inseps{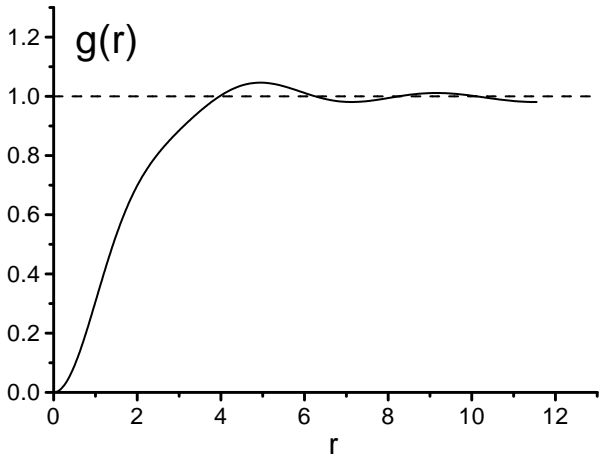}{0.35}
\caption{Pair correlation function of the $\nu=3/5$ parafermion state
for $N=18$ electrons on the sphere. $r$ is the great-circle distance.}
\label{fig:n0w0}
\end{figure}

In Fig.~\ref{fig:n0w0} we show the
pair correlation function for this same state but for the $N=18$ size system,
plotted as a LLL wavefunction, which describes the correlations of 
the guiding-center coordinates of the particles.
Again, the ``shoulder'' at small $r$ appears to be a characteristic of
states in which the particles form clusters, such as pairs, and is also
present for the Pfaffian state \cite{rr,jaingr}. 
In addition, the large-distance oscillations are strongly damped. 
This feature indicates that the state is incompressible.

We have also compared these states with the ground state obtained
by diagonalizing the Coulomb potential for the ${\cal N}=1$ Landau level.
In the LLL, the hierarchy states are most stable and
these new ones will not be competitive. The first excited Landau level
may be quite a different matter since $V_1$ is reduced compared to
$V_3$ (see e.g.\ Ref.\ \cite{hald}). Some typical values for  $V_1$, $V_3$,
and $V_5$ at $N_\phi=22$ are ${\cal N}=0$: 0.4681, 0.2998, 0.2422, and 
${\cal N}=1$: 0.4716, 0.3711, 0.2800. For the ${\cal N}=1$ Coulomb interaction
ground state, the hierarchy state has small overlap-squared, whereas 
parafermionic states have very large ones. (The hierarchy state was obtained 
as the ground state of a model pseudopotential consisting only of non-zero 
$V_1$. We caution that these two states occur at different values of $N_\phi$ 
because of the finite shifts on the sphere [$N_\phi=5N/3+1$ for the hierarchy].)
For the $k=3$ parafermion state we find 97\% for $N=15$ (where there
are 36 states in the $L=0$ Hilbert space) and 88\% for $N=18$ (319 $L=0$ 
states), compared to at best 1 or 2 \% for the hierarchy 3/5 state.
This is noteworthy since for these sizes we are very close to a Laughlin
2/3 state (i.e.\ the Laughlin 1/3 state of holes), because of the
finite shifts in the $N$--$N_\phi$ relations on the sphere. 
For $N=15$ the $N_\phi$  for our state coincides with that of a single
quasi-particle excitation of the $\nu=2/3$ fluid while for $N=18$
it is at the same flux as the $2/3$ condensate itself.
This is a clear disadvantage for the parafermion states
particularly if we try to vary the short-range component of the Haldane
pseudopotential $V_1$. Not surprisingly, increasing $V_1$ by a few
percent seems to favor the Laughlin state. However, one would expect the
hierarchy $\nu=3/5$ state, against which our state will be ultimately
competing for large sizes, to show a slower rate of stabilization upon
increasing $V_1$ away {}from its second Landau level value.

\begin{figure}
\inseps{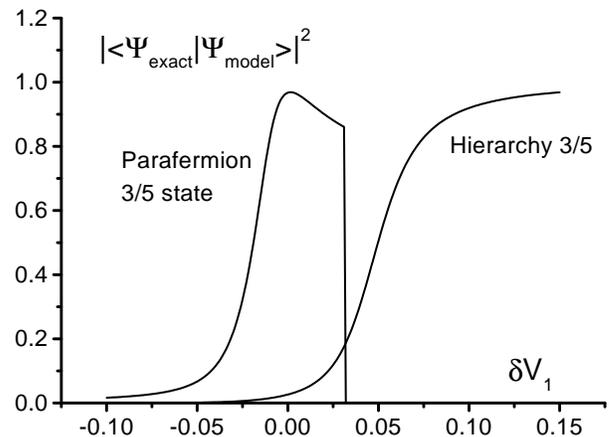}{0.35}
\caption{Overlap-squared of the two reference states, the $\nu=3/5$ parafermion
state ($N=15$) and the $\nu=3/5$ hierarchy state ($N=12$), with the
state obtained by diagonalizing the ${\cal N}=1$ Coulomb potential with
an added $\delta V_1$ component.}
\label{fig:n1w1}
\end{figure}

To study this issue we compare the overlap-squared of the ground state of 
$H=H_{{\rm Coul},{\cal N}=1}+\delta V_1$ with both our state and the hierarchy 
$3/5$ state. Fig.~\ref{fig:n1w1} shows these overlaps as a function of $\delta 
V_1$ for the $N=12$ (52 $L=0$ states) hierarchy and $N=15$ parafermion state. 
(The sharp drop in the parafermion curve at $\delta V_1$ about 0.03 is due to
a level crossing: for larger $\delta V_1$, the ground state has $L=2$.)
It appears that our state remains stable for increases of $\delta V_1$ of up 
to $7$--$8$\% {}from the Coulomb value. Note however that the hierarchy state 
is not fully stabilized until $V_1$ is increased by 20\% of its Coulomb value.  
For large sizes there will be less interference {}from $\nu=2/3$ and the 
stability domain for our state may extend well beyond 7--8\%.

\begin{figure}
\inseps{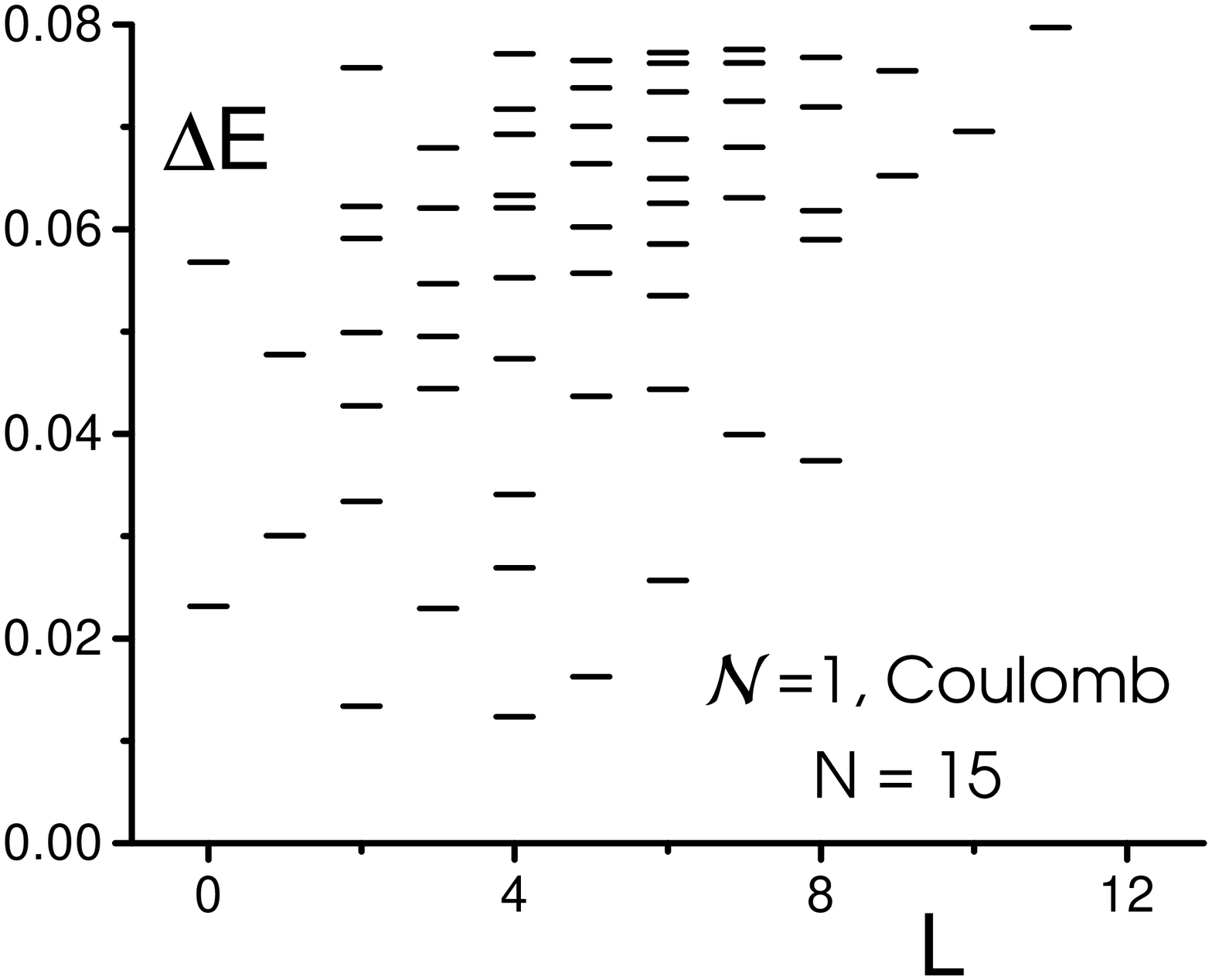}{0.35}
\caption{Same as Fig.~\protect\ref{fig:specM} but for the ${\cal N}=1$ Coulomb 
potential. The ground-state energy has been subtracted in the spectrum.}
\label{fig:n2w2}
\end{figure}

Finally, in Fig.~\ref{fig:n2w2} we show the low-lying spectrum for the pure 
Coulomb case for $N=15$. Again one finds some similarity with that of the 
four-body Hamiltonian (Fig.~\ref{fig:specM}), although in neither case is 
there a clear gap to a continuum of excited states. We defer precise gap 
estimations incorporating finite layer thickness and other effects that modify 
the gap values, as well as detailed studies of the quasiparticles, to future 
work.

%%%%%%%%%%%%%%%%%%%%%%%%%%%%%%%%%%%%%%%%%%%%%%%%%%
\section{Conclusion}
\label{conclusion}

To conclude, we have obtained the ground-state wavefunctions for the 
$k+1$-body Hamiltonians, and for the case $k=3$, we have counted quasihole 
states, and (for $M=1$) found excellent overlaps with the ground state 
of the Coulomb interaction for spin-polarized electrons in the first excited 
(${\cal N}=1$) LL, at total filling factor $\nu=2+3/5$; by a particle-hole 
transformation, this also applies at $\nu=2+2/5$. 

Our parafermion states contain clusters of $k$ particles. For the Pfaffian
state ($k=2$) at $\nu=1/2$, it has been suggested \cite{morf,rezhal} that it
may be favored in the ${\cal N}=1$ LL, because of this feature. That is, the 
correlation hole around each particle that is obtained by the Laughlin-style 
correlations, as a result of attaching $q$ vortices to each particle 
(for filling factor of the topmost LL equal to $1/q$) \cite{read94}, may not 
be sufficient to obtain the lowest energy in higher LLs, because of the 
``form factor'' associated with these LLs---the real space wavefunction 
does not vanish when particles coincide \cite{morf}. However, by forming a 
cluster of two or more particles, the energy gained {}from the larger 
correlation hole surrounding the cluster may outweigh that lost within the 
cluster itself. Very similar physics is suggested by recent work motivated by
the form of the pseudopotential interaction in higher LLs; the latter are also 
a consequence of the form factors. These works consider the formation of
crystalline phases in which the particles are clustered, so there is more than
one per unit cell \cite{shklov}. If this idea is correct, then liquids
containing clusters might also be expected to occur in higher LLs, possibly as
intermediate phases between the crystals and the usual hierarchy states (if we
consider varying the short-range part of the interaction away {}from its 
physical value, at fixed $\nu$). As the LL index ${\cal N}$ increases, larger 
clusters with $k\sim {\cal N}$ are expected to be favored \cite{shklov}. It is 
very interesting that recent experimental work \cite{jim} has observed 
highly-resistive, highly-anisotropic behavior, with nonlinear current-voltage 
characteristics, at total filling factors greater than 4, that is in the range 
${\cal N}=2$ to about $6$. This behavior, seen around the center of each LL 
for each spin component, appears consistent in principle with the notion of 
a uniaxial or ``striped'' crystal phase, which was also predicted in Ref.\ 
\cite{shklov} in this region (the triaxial crystals of clusters were predicted 
for $\nu$ away {}from $1/2$), and thus may support these physical pictures.  
There is clearly much still to be done to understand the physics in this
regime, in which we hope that the parafermion liquid states will play a role. 

%%%%%%%%%%%%%%%%%%%%%%%%%%%%%%%%%%%%%%%%%%%%%%%%%
\acknowledgements
We thank A.~Ludwig, G.~Moore, R.H.~Morf, and K.~Schoutens for helpful 
discussions. We also thank the Institute for Theoretical Physics, UCSB, program 
``Disorder and Interactions in Quantum Hall and Mesoscopic Systems'' 
for a stimulating environment in which this work was completed. This work was 
supported by NSF grants, nos.\ DMR-9157484 (NR), DMR-9420560 (ER), and at the 
ITP by NSF-PHY94-07194.   ER is also grateful to ITP for an ITP Scholar award.

%%%%%%%%%%%%%%%%%%%%%%%%%%%%%%%%%%%%%%%%%%%%%%%%%%
%\widetext
       
\widetext   

%%%%%%%%%%%%%%%%%%%%%%%%%%%%%%%%%%%%%%%%%%%%%%%%%%%

%\vspace*{\fill} 

\begin{references}
\bibitem{note}
   A sum over permutations within a triple $i$, $j$, $k$ is not
   required.
\bibitem{note2}
   In some cases studied in Ref.\ \cite{rr}, the number of zero-energy states
   was greater than the number of conformal blocks. In any case, the argument 
   given here does not establish completeness, so the number of conformal 
   blocks should be viewed as a lower bound on the degeneracy. However, 
   in the present case for general $k$ values, we expect the number of 
   zero energy states to saturate this bound.
\bibitem{schoutens}
   This result was pointed out to us by K.~Schoutens, private
   communication.
\end{references}
\end{document}